\documentclass[graybox]{svmult}
\usepackage{type1cm}
\usepackage{makeidx}
\usepackage{graphicx}
\usepackage{multicol}
\usepackage{here}
\usepackage[bottom]{footmisc}
\usepackage{newtxtext} 
\usepackage{newtxmath}
\usepackage{comment}
\usepackage{braket}
\usepackage{booktabs}
\usepackage{multirow}
\usepackage{colortbl}
\usepackage{tcolorbox}
\usepackage{lipsum}
\tcbuselibrary{breakable}
\usepackage{ascmac}
\usepackage{appendix}
\usepackage{subfigure}
\makeindex

\newcommand{\erfc}{{\mathrm{erfc}}}
\newcommand{\igamc}{{\mathrm{igamc}}}

\begin{document}

\title*{Quantum Random Numbers Generated by a Cloud Superconducting Quantum Computer}
\titlerunning{Quantum Random Numbers Generated by a Cloud Superconducting Quantum Comp.}
\author{Kentaro Tamura and Yutaka Shikano}
\institute{Kentaro Tamura \at Department of Applied Physics and Physico-Informatics, Keio University, 3-14-1 Hiyoshi, Kohoku, Yokohama, 223-8522 Japan, \email{cicero@keio.jp}
\and Yutaka Shikano \at Quantum Computing Center, Keio University, 3-14-1 Hiyoshi, Kohoku, Yokohama, 223-8522 Japan \email{yutaka.shikano@keio.jp} and Institute for Quantum Studies, Chapman University, 1 University Dr., Orange, CA 92866 USA}
\maketitle

\abstract{
A cloud quantum computer is similar to a random number generator in that its physical mechanism is inaccessible to its users. In this respect, a cloud quantum computer is a black box. In both devices, its users decide the device condition from the output. A framework to achieve this exists in the field of random number generation in the form of statistical tests for random number generators. In the present study, we generated random numbers on a 20-qubit cloud quantum computer and evaluated the condition and stability of its qubits using statistical tests for random number generators. As a result, we observed that some qubits were more biased than others. Statistical tests for random number generators may provide a simple indicator of qubit condition and stability, enabling users to decide for themselves which qubits inside a cloud quantum computer to use.}

\section{Introduction}
\label{sec:1}
Given a coin with an unknown probability distribution, there are two approaches to decide whether the coin is fair~\cite{Tamura}. The first approach is to examine the coin itself; one expects an evenly shaped coin to yield fair results. The second approach is to actually toss the coin a number of times to see if the output is sound. In this approach, the coin is treated as a black box. 
A random number generator is similar to a coin in that it is expected to produce unbiased and independent 0s and 1s. Unlike a coin, however, the physical mechanism of a random number generator is often inaccessible to its users. Therefore, users rely on statistical tests to decide the fairness of the device from its output. 

Random number generators play an important role in cryptography, particularly in the context of key generation. For example, the security of the RSA cryptosystem is based on keys that are determined by random choices of two large prime numbers~\cite{rsa}. If the choices of prime numbers are not random, an adversary could predict future keys and hence compromise the security of the system. Randomness in cryptography derives from what is called the seed. The seed is provided by physical random number generators~\cite{seed1,seed2}. It is required that the physical mechanism of a physical random number generator remains a black box for the seed to be unpredictable. Given that the measurement outcomes are theoretically unpredictable in quantum mechanics, random number generators based on quantum phenomena are a promising source of unpredictability~\cite{bell, qrng, Herrero-Collantes2016}.

Cloud quantum computers are quantum computers that are accessed online~\cite{cloud, news, news2, nmrcloud, opticalcloud, QCReview}. In order to use a cloud quantum computer, users are required to send programs specifying the quantum circuit to be executed and the number of times the circuit should be run~\cite{software}. When a user's turn arrives, the quantum computer executes the program and returns the results~\cite{preskill}. A similarity between random number generators and cloud quantum computers is that its users do not have direct access to the physical mechanism of the device. So, as far as the users are concerned, both random number generators and cloud quantum computers are black boxes. In the field of random number generation, much research has been done on how to characterize the device from its output. This lead to the creation of statistical tests for random number generators. The present study aims to introduce the idea of statistical tests for random number generators to the field of cloud quantum computing. This aim is supported by three points. Firstly, the cloud quantum computer is a black box to its users, which is also the case with random number generators. Secondly, quantum computers become random number generators when given certain programs. Finally, the cloud quantum computer lacks a simple benchmark that would enable its users to decide the condition of the device. 

The rest of this article is organized as follows. In Section 2, statistical tests for random number generators is generally explained. In Section 3, a group of statistical tests called the NIST SP 800-22 is reviewed. In Section 4, we present the results of the statistical analysis of random number samples obtained from the cloud quantum computer, IBM 20Q Poughkeepsie, and the test results of the eight statistical tests from the NIST SP 800-22. Finally, Section 5 is devoted to the conclusion. In the appendix, a measure of uniformity often employed in the field of cryptography,  the min-entropy, is explained. 

\section{Statistical Tests for Random Number Generators}
\label{sec:2}
Statistical tests for random number generators are necessary to confirm that a random number generator is suitable for use in encryption processes~\cite{stat review}. Random number generators used in this context are required to have unpredictability. This means that given any subset of a sequence produced by the device, no adversary can predict the rest of the sequence, including the output from the past. Statistical tests aim to detect random number generators that produce sequences with a significant bias and/or correlation. 

When subjected to statistical tests, a random number generator is considered a black box. This means that the only information available is its output. Under the null hypothesis that the generator is unbiased and independent, one expects its output to have certain characteristics. The characteristics of the output are quantified by the test statistic, whose probability distribution is known. From the test statistic, the probability that a true random number generator produces an output with a worse test statistic value is calculated. This probability is called the p-value. If the p-value is below the level of significance $\alpha$, the generator fails the test and the null hypothesis that the generator is unbiased and independent is rejected. Since statistical tests for random number generators merely rule out significantly biased and/or correlated generators, these tests do not verify that a device is the ideal random number generator. Nevertheless, a generator that passes the tests is more reliable than a generator that doesn't. This is why statistical tests are usually organized in the form of test suites, so as to be comprehensive. Some well known test suites are the NIST SP 800-22~\cite{nist}, TestU01~\cite{TestU01}, and the Dieharder test. 

Because statistical tests are designed to check for statistical anomalies under the hypothesis that the generator is unbiased, a biased random number generator would naturally fail the tests. This can be a problem when testing quantum random number generators, as they can be biased and unpredictable at the same time. Given that statistically faulty generators can still be unpredictable, the framework of statistical tests fails to capture the essence of randomness: unpredictability. There have been attempts to assure the presence of unpredictability by exploiting quantum inequalities, but they have not reached the point of replacing statistical tests altogether.

\section{NIST SP 800-22}
The NIST SP 800-22 is a series of statistical tests for cryptographic random number generators provided by the National Institute of Standards and Technology~\cite{nist}. Random number generators for cryptographic purposes are required to have unpredictability, which is not strictly necessary in other applications such as simulation and modeling, but is a crucial element of randomness. The test suite contains 16 tests, each test with a different test statistic to characterize deviations of binary sequences from randomness. The entire testing procedure of the NIST SP 800-22 is divided into 3 steps. The first step is to subject all samples to the 16 tests. For each sample, each test returns the probability that the sample is obtained from an unbiased and independent RNG. This probability, which is called the p-value, is then compared to the level of significance $\alpha = 0.01$. If the p-value is under the level of significance, the sample fails the test. The second step involves the proportion of passed samples for each test. Under the level of significance $\alpha = 0.01$, 1\% of samples obtained from an unbiased and independent RNG is expected to fail each test. If the proportion of passed samples is too high or too low, the RNG fails the test. Finally, p-value uniformity is checked for each test. Suppose one tested 100 binary samples. This yields 100 p-values per test. If the samples are independent, the p-values should be uniformly distributed for all tests. The distribution of p-values is checked via the chi-squared test.

In the following sections, 8 tests from the NIST SP 800-22 are explained. The input sequence will be denoted $\varepsilon = \varepsilon_1, \varepsilon_2, \cdots, \varepsilon_n$, and the $i$th element $\varepsilon_i$.
\begin{table}[H]
\caption{The minimum length $n$ required for each test in order to obtain meaningful results. The tests not employed in the present study are shaded in grey. Note that the tests will be referred to by their test \# in Sec.~\ref{sec:3}.}
\label{tab:min_sample}
\centering
\begin{tabular}{lll}\hline
Test \#                    & Test name                  & Minimum length \\ \hline \hline
1 & Frequency  & $n\geq 100$   \\
2 & Frequency within a block  & $n\geq100$\\
3 & Runs                      & $n\geq100$\\
4 & Longest run of ones       & $n\geq 128$\\
\rowcolor[rgb]{0.9, 0.9, 0.9} & Binary matrix rank     & $n\geq 38912$\\
5 & DFT   & $n\geq 1000$\\
\rowcolor[rgb]{0.9, 0.9, 0.9} &Non-overlapping T. M.    & $n\geq 8m - 8$\\
\rowcolor[rgb]{0.9, 0.9, 0.9} & Overlapping T. M.       & $n\geq 10^6$\\
\rowcolor[rgb]{0.9, 0.9, 0.9} &Maurer's Universal       & $n > 387840$\\
\rowcolor[rgb]{0.9, 0.9, 0.9} &Linear complexity        & $n\geq 10^6$\\
\rowcolor[rgb]{0.9, 0.9, 0.9} &Serial                   & $n > 16$    \\
6 & Approximate entropy       & $n > 64$   \\
7 & Cumulative sums (forward)           & $n\geq 100$ \\
8 & Cumulative sums (backward)          & $n\geq 100$ \\
\rowcolor[rgb]{0.9, 0.9, 0.9} &Random excursions         & $n\geq 10^6$ \\
\rowcolor[rgb]{0.9, 0.9, 0.9} &Random excursions variant & $n\geq 10^6$ \\ \hline
\end{tabular}
\end{table}
\subsection{Frequency Test}
The frequency test aims to test whether a sequence contains a reasonable proportion of 0s and 1s. If the probability of obtaining the sequence from an independent and unbiased random number generator is lower than 1 \%, it follows that the random number generator is not ``independent and unbiased". The minimum sample length required for this test is 100. 
\begin{tcolorbox}[colback=white!5!white,colframe=black!5!black,enforce breakable,pad at break*=1mm,title=Test Description]
  \begin{enumerate}
    \item Convert the sequence into $\pm{1}$ using the formula:
    $X_i = 2\varepsilon_i-1$.
    \item Add the elements of $X$ together to obtain $S_n$.
    \item Compute test statistic: $s_{{\mathrm{obs}}} = |S_n|/\sqrt{n}$.
    \item Compute p-value $=\erfc(s_{{\mathrm{obs}}} / \sqrt{2})$ using complementary error function shown as 
    \begin{align}
    \erfc(z) = \frac{2}{\sqrt{\pi}}\int_z^\infty e^{-u^2}du.
    \label{eq:erfc}
    \end{align}
    \item Compare p-value to 0.01. If p-value $\geq$ 0.01, then the sequence passes the test. Otherwise, the sequence fails.
  \end{enumerate}
\end{tcolorbox}
\begin{tcolorbox}[colback=white,enforce breakable,pad at break*=1mm]
Example: $\varepsilon = 1001100010$, length $n = 10$.
    \begin{enumerate}
        \item $1,0,0,1,1,0,0,0,1,0 \rightarrow +1,-1,-1,+1,+1,-1,-1,-1,+1,-1$.
        \item $S_{10} = 1-1-1+1+1-1-1-1+1-1 = -2$.
        \item $s_{{\mathrm{obs}}} = |-2|/\sqrt{10} \approx 0.632455$.
        \item P-value $=\erfc(s_{\mathrm{obs}} / \sqrt{2}) \approx 0.527089$.
        \item P-value $=0.527089 > 0.01 \rightarrow $ the sequence passes the test.
    \end{enumerate}
\end{tcolorbox}
This test is equivalent to testing the histogram for bias. Because the test only considers the proportion of 1s, sequences such as $0000011111$ or $0101010101$ would pass the test. Failing this test means that the sample is overall biased.
\subsection{Frequency Test Within a Block}
Firstly, the sequence is divided into $N$ blocks of size $M$. The frequency test is then applied to the respective blocks. As a result, one obtains $N$ p-values. The second part of this test aims to check whether the variance of the p-values is by chance or not. This is called the chi-squared ($\chi^2$) test. For meaningful results, a sample with a length of at least 100 is required. The following is the test description.
\begin{tcolorbox}[colback=white!5!white,colframe=black!5!black,enforce breakable,pad at break*=1mm,title=Test Description]
  \begin{enumerate}
    \item Divide the sequence into $N = \lfloor \frac{n}{M} \rfloor$ non-overlapping blocks of size $M$. 
    \item Determine the proportion of 1s in each block using 
    \begin{align}
      \pi_i = \frac{\sum_{j=1}^M \varepsilon_{(i-1)M+j}}{M}.
      \label{eq:pro_of_1s}
    \end{align}
    \item Compute $\chi^2$ statistic $\chi^2_{{\mathrm{obs}}} = 4M\sum_{i=1}^N \left(\pi_i-\frac{1}{2} \right)^2$. 
    \item Compute p-value $= 1 - \igamc \left(\frac{N}{2}, \frac{\chi^2_{{\mathrm{obs}}}}{2} \right)$. Note that $\igamc$ stands for the incomplete gamma function.
    \begin{align}
      \Gamma(z) &= \int_0^\infty t^{z-1} e^{-t}\\
      \igamc(a,x) &\equiv \frac{1}{\Gamma(a)}\int_0^x e^{-t} t^{(a-1)}dt
      \label{eq:igamc}
    \end{align}
    \item Compare p-value to 0.01. If p-value $\geq$ 0.01, then the sequence passes the test. Otherwise, the sequence fails.
  \end{enumerate}
\end{tcolorbox}
\begin{tcolorbox}[colback=white,enforce breakable,pad at break*=1mm]
Example: $\varepsilon = 1001100010$, length: $n = 10$.
    \begin{enumerate}
        \item If $M=3$, then $N=3$ and the blocks are $100$, $110$, $001$. The final 0 is discarded.
        \item $\pi_1 = 1/3$, $\pi_2 = 2/3$, $\pi_3 = 1/3$.
        \item $\chi^2_{{\mathrm{obs}}} = 4M\sum_{i=1}^N(\pi_i-\frac{1}{2})^2$.
        \item $\chi^2_{{\mathrm{obs}}} = 4\times 3 \times \left\{
    (\frac{1}{3}-\frac{1}{2})^2+ (\frac{2}{3}-\frac{1}{2})^2+ (\frac{1}{3}-\frac{1}{2})^2
    \right\} = 1$.
        \item P-value $=1 - \igamc(\frac{3}{2}, \frac{1}{2})=0.801252$.
        \item P-value $=0.801252 > 0.01 \rightarrow$ the sequence is passes the test.
    \end{enumerate}
\end{tcolorbox}
This test divides the sequence into blocks and checks each block for bias. Depending on the block size, samples such as $001100110011$ or $101010101010$ could pass the test. Failing this test means that certain sections of the sequence are biased.
\subsection{Runs Test}
The proportion of 0s and 1s does not suffice to identify a random sequence. A run, which is an uninterrupted sequence of identical bits, is also a factor to be taken into account. The runs test determines whether the lengths and oscillation of runs in a sequence is as expected from a random sequence. A minimum sample length of 100 is required for this test. The following is the test description.
\begin{tcolorbox}[colback=white!5!white,colframe=black!5!black,enforce breakable,pad at break*=1mm,title=Test Description]
\begin{enumerate}
  \item Compute proportion of ones $\pi = \left( \sum_j \varepsilon_j \right) / n$.
  \item If the sequence passes frequency test, proceed to next step. Otherwise, the p-value of this test is 0.
  \item Compute test statistic $V_n({\mathrm{obs}}) = \sum_{k=1}^{n-1}(\varepsilon_k\oplus \varepsilon_{k+1})+1$, where $\oplus$ stands for the XOR operation.
  \item Compute p-value $= \erfc\left(\frac{|V_n({\mathrm{obs}})-2n\pi(1-\pi)|}{2\sqrt{2n}\pi(1-\pi)} \right)$.
  \item Compare p-value to 0.01. If p-value $\geq$ 0.01, then the sequence passes the test. Otherwise, the sequence fails.\\
\end{enumerate}
\end{tcolorbox}
\begin{tcolorbox}[colback=white,enforce breakable,pad at break*=1mm]
Example: $\varepsilon = 1010110001$, length $n = 10$.
    \begin{enumerate}
        \item $\pi = \frac{5}{10} = 0.5$.
        \item $|\pi - 0.5| = 0 < \frac{2}{\sqrt{n}} =  \frac{2}{\sqrt{10}} = 0.63\rightarrow$ test is applicable.
        \item $V_{10}({\mathrm{obs}}) = (1 + 1 + 1 + 1 + 0 + 1 + 0 + 0+ 1) + 1 = 7$.
        \item P-value $= \erfc\left(\frac{|7-2\times 10\times 0.5\times (1-0.5)|}{2\times \sqrt{2\times 10}\times 0.5 \times(1-0.5)} \right) = 0.21$.
        \item P-value $= 0.21\geq 0.01$, so sequence passes the test.
    \end{enumerate}
\end{tcolorbox}
\subsection{The Longest Run of Ones Within a Block Test}
This test determines whether the longest runs of ones $111\cdots$ within blocks of size M is consistent with what would be expected in a random sequence. The possible values of M for this test are limited to three values, namely, 8, 128 and 10,000, depending on the length of the sequence to be tested. 
\begin{tcolorbox}[colback=white!5!white,colframe=black!5!black,enforce breakable,pad at break*=1mm,title=Test Description]
\begin{enumerate}
  \item Divide the sequence into blocks of size M.
  The choices of M and N are determined in regard to the length of the sequence.
  N denotes the number of blocks, and the elements exceeding the number of blocks are discarded.
  \begin{table}[H]
  \centering
  \caption{Choices of M for the longest runs of ones within a block test.}
  \label{tab:longest_runs_of_1s}
  \begin{tabular}{|l|l|} \hline
  Minimum length $n$ & $M$ \\ \hline \hline
  128 & 8 \\ \hline
  \textcolor{red}{6272} & \textcolor{red}{128} \\ \hline
  750,000 & 10000 \\ \hline
  \end{tabular}
  \end{table}
  \item Classify each block into the following categories regarding $M$ and
  the length of the longest run in each block. See Table~\ref{tab:longest_runs_of_1s_2}.
  \begin{table}[H]
  \centering
  \caption{Classifications of each block.}
  \label{tab:longest_runs_of_1s_2}
  \begin{tabular}{|c|c|c|c|} \hline
  Classes $v_i$ & $M \geq 8$ & $M \geq 128$ & $M \geq 100000$\\ \hline \hline
  $v_0$ & $\leq 1$ & $\leq 4$ & $\leq 10$ \\ \hline
  $v_1$ & 2 & 5 & 11\\ \hline
  $v_2$ & 3 & 6 & 12\\ \hline
  $v_3$ & $\geq 4$ & 7 & 13\\ \hline
  $v_4$ & & 8 & 14\\ \hline
  $v_5$ & & $\geq 9$ & 15\\ \hline
  $v_6$ & & & $\geq 16$\\ \hline
  \end{tabular}
  \end{table}
  \item Compute $\chi^2({\mathrm{obs}}) = \sum_{i = 0}^{K}\frac{(v_i - N\pi_i)^2}{N\pi_i}$. Note that $K$, $N$ and $\pi_i$ are determined by $M$. See Tables~\ref{tab:longest_runs_of_1s_3} and \ref{tab:longest_runs_of_1s_4}.
  \begin{table}[H]
  \centering
  \caption{Values of $K$ and $N$ corresponding to $M$.}
  \label{tab:longest_runs_of_1s_3}
  \begin{tabular}{|l|l|l|} \hline
  $M$ & $K$ & $N$\\ \hline \hline
  8 & 3 & 16\\ \hline
  128 & 5 & 49\\ \hline
  10000 & 6 & 75\\ \hline
  \end{tabular}
  \end{table}
  \begin{table}[H]
  \centering
  \caption{Values of $\pi_i$ corresponding to $K$ and $M$.}
  \label{tab:longest_runs_of_1s_4}
  \begin{tabular}{|c|c|c|c|} \hline
    & \multicolumn{3}{|c|}{$\pi_i$}\\ \hline
    Classes & $K = 3$, $M = 8$ &$K = 5$, $M = 128$ &$K = 6$, $M = 10000$\\ \hline \hline
  $v_0$ & 0.2148 & 0.1174 & 0.0882\\ \hline
  $v_1$ & 0.3672 & 0.2430 & 0.2092\\ \hline
  $v_2$ & 0.2305 & 0.2493 & 0.2483\\ \hline
  $v_3$ & 0.1875 & 0.1752 & 0.1933\\ \hline
  $v_4$ &  & 0.1027 & 0.1208\\ \hline
  $v_5$ &  & 0.1124 & 0.0675\\ \hline
  $v_6$ &  &  & 0.0727\\ \hline
  \end{tabular}
  \end{table}
  \item Compute p-value $= 1 - \igamc\left(\frac{K}{2},\frac{\chi^2({\mathrm{obs}})}{2}\right)$.
  \item Compare p-value to 0.01. If p-value $\geq$ 0.01, then the sequence passes the test. Otherwise, the sequence fails.
\end{enumerate}
\end{tcolorbox}
\begin{tcolorbox}[colback=white,enforce breakable,pad at break*=1mm]
Example: $n = 10000$
    \begin{enumerate}
        \item $M = 128$ and $N = 49$. The remaining $3728$ elements are discarded. 
        \item The counts for the longest run of ones are $v_0 = 6$, $v_1 = 10$, $v_2 = 10$ , $v_3 = 7$, $v_4 = 7$, and $v_5 = 9$.
        \item
        \begin{align*}
            \chi^2({\mathrm{obs}}) &　= \frac{(6 - 49\times 0.1174)^2}{49\times 0.1174} + \frac{(10 - 49\times 0.2430)^2}{49\times 0.2430} \\&　+ \frac{(10 - 49\times 0.2493)^2}{49\times 0.2493} + \frac{(7 - 49\times 0.1752)^2}{49\times 0.1752}　\\&　+ \frac{(7 - 49\times 0.1027)^2}{49\times 0.1027} + \frac{(9 - 49\times 0.1124)^2}{49\times 0.1124}\\
            & = 3.994459.
            \end{align*}
        \item P-value $= 1 - \igamc\left(\frac{5}{2},\frac{3.994459}{2}\right) = 0.550214$.
        \item P-value $= 0.550214 \geq 0.01$, so the sequence passes the test.
    \end{enumerate}
\end{tcolorbox}

\subsection{Discrete Fourier Transform Test}
This test checks for periodic patterns in the sequence by performing a discrete Fourier transform (DFT). The minimum sample length required for this test is 1000. The following is the test description.
\begin{tcolorbox}[colback=white!5!white,colframe=black!5!black,enforce breakable,pad at break*=1mm,title=Test Description]
\begin{enumerate}
  \item Convert the sequence $\varepsilon$ of 0s and 1s into a sequence $X$ of $-1$s and $+1$s.
  \item Apply a DFT on X: $S = DFT(X)$. This should yield a sequence of complex variables representing the periodic components
  of the sequence of bits at different frequencies.
  \item Compute $M = {\mathrm{modulus}}(S') \equiv |S'|$, where $S'$ is the first $\frac{n}{2}$ elements of $S$. This produces a sequence of peak heights.
  \item Compute $T = \sqrt{\left(\log_e\frac{1}{0.05}\right)}$. This is the 95 \% peak height threshold value. 95 \% of the values obtained by the test should not exceed $T$ for a random sequence.
  \item Compute $N({{\mathrm{ideal}}}) = \frac{0.95n}{2}$,
  which is the expected theoretical number of peaks that are less than $T$.
  \item Compute $N({\mathrm{obs}})$, which is the actual number of peaks in $M$ that are less than $T$.
  \item Compute $d = \frac{N({{\mathrm{ideal}}})-N({\mathrm{obs}})}{\sqrt{n\cdot 0.95\cdot 0.05\cdot \frac{1}{4}}}$.
  \item Compute p-value $= \erfc\left(\frac{|d|}{\sqrt{2}}\right)$.
  \item Compare p-value to 0.01. If p-value $\geq$ 0.01, then the sequence is passes the test. Otherwise, the sequence fails.
\end{enumerate}
\end{tcolorbox}
This test checks for periodic features. Samples with periodic features may look like $0110011001100110$ or $010010100101001$ among various other possibilities. Failing this test suggests that the sample has periodic patterns. It is noted that the probability distribution of the test statistic $d$ should be rectified as it does not converge to the standard normal distribution~\cite{hamano}.
\begin{tcolorbox}[colback=white,enforce breakable,pad at break*=1mm]
Example: $\varepsilon = 1001010011$, length $n = 10$.
    \begin{enumerate}
        \item $X = 2\varepsilon_1 - 1, 2\varepsilon_2 - 1, \ldots, 2\varepsilon_n - 1= 1,-1,-1,1,-1,1,-1,-1,1,1$.
        \item $N({{\mathrm{ideal}}}) = 4.75$.
        \item $N({\mathrm{obs}}) = 4$.
        \item $d = \frac{(4.75-4)}{\sqrt{10\cdot 0.95\cdot 0.05\cdot \frac{1}{4}}} = 2.147410$.
        \item P-value $= \erfc\left(\frac{|2.147410|}{\sqrt{2}}\right) = 0.031761$.
        \item P-value $= 0.031761 \geq 0.01$, so the sequence passes the test.
    \end{enumerate}
\end{tcolorbox}
\subsection{Approximate Entropy Test}
The approximate entropy test compares the frequency of $m$-bit overlapping patterns with that of $(m+1)$-bit patterns in the sequence. It checks whether the relation of two frequencies is what is expected from an unbiased and independent RNG. The level of significance is $\alpha = 0.01$. This test can be applied to samples with lengths equal to or larger than 64. The test description is below.
\begin{tcolorbox}[colback=white!5!white,colframe=black!5!black,enforce breakable,pad at break*=1mm,title=Test Description]
\begin{enumerate}
    \item Append the first $m-1$ bits of the sequence to the end of the sequence.
    \item Divide the sequence into overlapping blocks with a length of $m$.
    \item There are $2^m$ possible m-bit blocks. Count how many of each possible block there are in the sequence.
    \item Compute $\frac{\mathrm{count}}{n}\log_e(\frac{\mathrm{count}}{n})$ for each count.
    \item Compute the sum of all counts $\varphi_m$.
    \item Replace $m$ with $m+1$ and repeat steps 1 through 5 to obtain $\varphi_{m+1}$.
    \item Calculate test statistic $\mathrm{obs} = 2n(\log_e(n) - (\varphi_m - \varphi_{m+1}))$.
    \item Derive p-value $= 1 - \igamc(2^{(m-1)}, \mathrm{obs}/2)$.
    \item Compare p-value with level of significance $\alpha = 0.01$. If p-value $\geq 0.01$, the result is pass. Otherwise, the sequence fails the test.
\end{enumerate}
\end{tcolorbox}
\begin{tcolorbox}[colback=white,enforce breakable,pad at break*=1mm]
Example: $\varepsilon = 1011010010$, length $n = 10$, $m = 3$.
    \begin{enumerate}
        \item $\varepsilon = \textcolor{red}{10}11010010$ $\rightarrow$ $1011010010\textcolor{red}{10}$.
        \item $101101001010$ $\rightarrow$ $101, 011, 110, 101, 010, 100, 001, 010, 101, 010$.
        \item $"000": 0,"001": 1,\\"010": 3,"011": 1,"100": 1,"101": 3,"110": 1 ,"111": 0$.
        \item $"000": 0, "001": 0.1\log_e (0.1), "010": 0.3 \log_e(0.3),\\ "011": 0.1\log_e(0.1), "100": 0.1\log_e(0.1),
    "101": 0.3\log_e(0.3),\\ "110": 0.1\log_e(0.1), "111": 0$.
        \item $\varphi_3 = -1.643418$
        \item $\varphi_{3+1} = -2.025326$.
        \item $\mathrm{obs} = 2\times 10\times (\log_e(10) - (-1.643418 - (-2.025326))) = 6.224774$.
        \item P-value $= 1 - \igamc(2^{(3-1)}, \mathrm{6.224774}/2) = 0.622069$.
        \item P-value $= 0.622069 \geq 0.01$. The sequence passes the test.
    \end{enumerate}
\end{tcolorbox}
The approximate entropy test checks for correlation between the number of $m$-bit patterns and $(m+1)$-bit patterns in the sequence. The difference between the number of possible $m$-bit patterns and the number of possible $(m+1)$-bit patterns in the sequence is computed, and if this difference is too small or too large, the two patterns are correlated. 
\subsection{Cumulative Sums Test}
The cumulative sums test is basically a random walk test. It checks how far from 0 the sum of the sequence in terms of $\pm 1$ reaches. For a sequence that contains uniform and independent 0s and 1s, the sum should be close to 0. This test requires a minimum sample length of $100$.
\begin{tcolorbox}[colback=white!5!white,colframe=black!5!black,enforce breakable,pad at break*=1mm,title=Test Description]
\begin{enumerate}
    \item Convert 0 to -1 and 1 to +1.
    \item In forward mode, compute the sum of the first  $i$ elements of $X$. In backward mode, compute the sum of the last $i$ elements of $X$.
    \item Find the maximum value $z$ of the sums.
    \item Compute the following p-value. $\Phi$ is the cumulative distribution function for the standard normal distribution.
    \begin{align}
        \mathrm{P\mathchar`-value} = 1 - &\sum_{k = \left(\frac{-n}{z} + 1 \right)/4}^{\left(\frac{n}{z} - 1\right)/4}
        \left[
        \Phi
        \left(
        \frac{(4k+1)z}{\sqrt{n}}
        \right)
        - \Phi
        \left(
        \frac{(4k-1)z}{\sqrt{n}}
        \right)
        \right] \nonumber \\
        +&\sum_{k = \left(\frac{-n}{z} - 3 \right)/4}^{\left(\frac{n}{z} - 1 \right)/4}
        \left[
        \Phi
        \left(
        \frac{(4k+3)z}{\sqrt{n}}
        \right)
        - \Phi
        \left(
        \frac{(4k+1)z}{\sqrt{n}}
        \right)
        \right].
        \label{eq:culminative}
    \end{align}
    \item Compare p-value to $\alpha = 0.01$. If p-value $\geq 0.01$, the result is pass. Otherwise, the sequence fails the test.
\end{enumerate}
\end{tcolorbox}
\begin{tcolorbox}[colback=white,enforce breakable,pad at break*=1mm]
Example: $\varepsilon = 1011010010$, length $n = 10$.
    \begin{enumerate}
        \item $\varepsilon = 1011010010$ $\rightarrow$ $X = 1, -1, 1, 1, -1, 1, -1, -1, 1, -1$.
        \item Forward mode: $S_1 = 1$, $S_2 = 1 + (-1) = 0$, $S_3 = 1 + (-1) + 1 = 2$,\\ $S_4 = 1 + (-1) + 1 + 1$, $S_5 = 1 + (-1) + 1 + 1 + (-1) = 1$,\\ $S_6 = 1 + (-1) + 1 + 1 + (-1) + 1 = 2$, $S_7 = 1 + (-1) + 1 + 1 + (-1) + 1 + (-1) = 1$, $S_8 = 1 + (-1) + 1 + 1 + (-1) + 1 + (-1) + 1 = 2$,\\ $S_9 = 1 + (-1) + 1 + 1 + (-1) + 1 + (-1) + 1 + (-1) = 1$.
        \item In forward mode, the maximum value is $z = 2$.
        \item P-value $= 0.941740$ for both forward and backward.
        \item P-value $= 0.941740 \geq 0.01$. The sequence passes the test.
    \end{enumerate}
\end{tcolorbox}
Once the p-value has been calculated for all tests and samples, the proportion of samples that passed the test is computed for each test. Let us consider a case where 1000 samples were subjected to each of the 15 tests. This results in $1000$ p-values per test. For example, if $950$ out of $1000$ samples passed the frequency test, the proportion of passed samples is $0.95$. If the proportion of passed samples falls within the following range for all 15 tests, the samples pass the second step of the NIST SP 800-22. The acceptable range of proportion is calculated with 
\begin{align}
    (1-\alpha)\pm 3\sqrt{\frac{\alpha(1-\alpha)}{m}},
    \label{eq:pro_range}
\end{align}
where $\alpha$ stands for the level of significance and $m$ the sample size. It is noted that it is controversial whether the coefficient should be 3. A suggestion that the coefficient should be 2.6 exists \cite{nist_review}. In the case of the current example, Eq. (\ref{eq:pro_range}) can be calculated using $\alpha = 0.01$ and $m = 1000$ as
\begin{align}
    (1-0.01)\pm 3\sqrt{\frac{0.01(1-0.01)}{1000}} = 0.99 \pm  0.0094.
    \label{eq:pro_range_ex}
\end{align}
From the fact that $0.95$ is not within the acceptable range, it follows that the samples fail the frequency test. The same process is done with all 16 tests, and unless the samples pass all tests, the result is that the hypothesis that the RNG is unbiased and independent is rejected. 

The final step of the NIST SP 800-22 is to evaluate the p-value uniformity of each test. In order to perform the chi-squared ($\chi^2$) test, the p-value is divided into 10 regions: $[k,k+0.1)$ for $k = 0, 1, \ldots, 9$. The test statistic is given by
\begin{align}
    \chi^2 = \sum_{i=1}^{10}\frac{(\mathrm{number\ of\ samples\ in\ \mathit{i}-th\  region}-\mathrm{sample\ size}/10)^2}{\mathrm{sample\ size}/10}.
    \label{eq:chi_teststat}
\end{align}
When the number of samples in each region is 2, 8, 10, 13, 17, 17, 13, 10, 8, 2, the test statistic \ref{eq:chi_teststat} is calculated as $\chi^2 = 25.200000$. From $\chi^2$, the p-value is
\begin{align}
    \mathrm{p\mathchar`-value} = \igamc
    \left(
    \frac{9}{2}, \frac{\chi^2}{2}
    \right).
    \label{eq: p_uniformity}
\end{align}
Therefore, in the current example where $\chi^2 = 25.200000$, the p-value is $0.002758$. The level of significance for the p-value uniformity is $\alpha = 0.0001$. So when the p-value is $0.002758$, it follows that the p-value distribution is uniform. The p-value uniformity test requires at least $55$ samples. As mentioned before, it is remarked that passing the NIST SP 800-22 does not ensure a sequence to be truly random~\cite{nist_cri1, nist_cri2, nist_cri3}.

\section{Quantum Random Number Generation on the Cloud Quantum Computer}
\label{sec:3}
According to quantum mechanics, the measurement outcomes of the superposition state $(\ket{0}+\ket{1})/\sqrt{2}$ along the computational basis ideally form random number sequences. This means that the resulting sequences are expected to pass the statistical tests for RNGs explained previously. Here, the computational basis, $\ket{0}$ and $\ket{1}$, spans the two-dimensional Hilbert space. In a quantum computer, the desired state $(\ket{0}+\ket{1})/\sqrt{2}$ is generated from the initial state $\ket{0}$ by applying the Hadamard gate to a single quantum bit (qubit). Note that in this process, the initial state is always the same. Unlike classical random number generators and pseudorandom number generators that require random seeds to produce independent sequences, quantum random number generators are capable of producing independent sequences with the same seed. This reduces the risk of the output of a random number generator being predicted from the seed, because all possible outputs come from the same seed.

\begin{figure}[ht]
 \centering
  \includegraphics[width=0.9\textwidth]{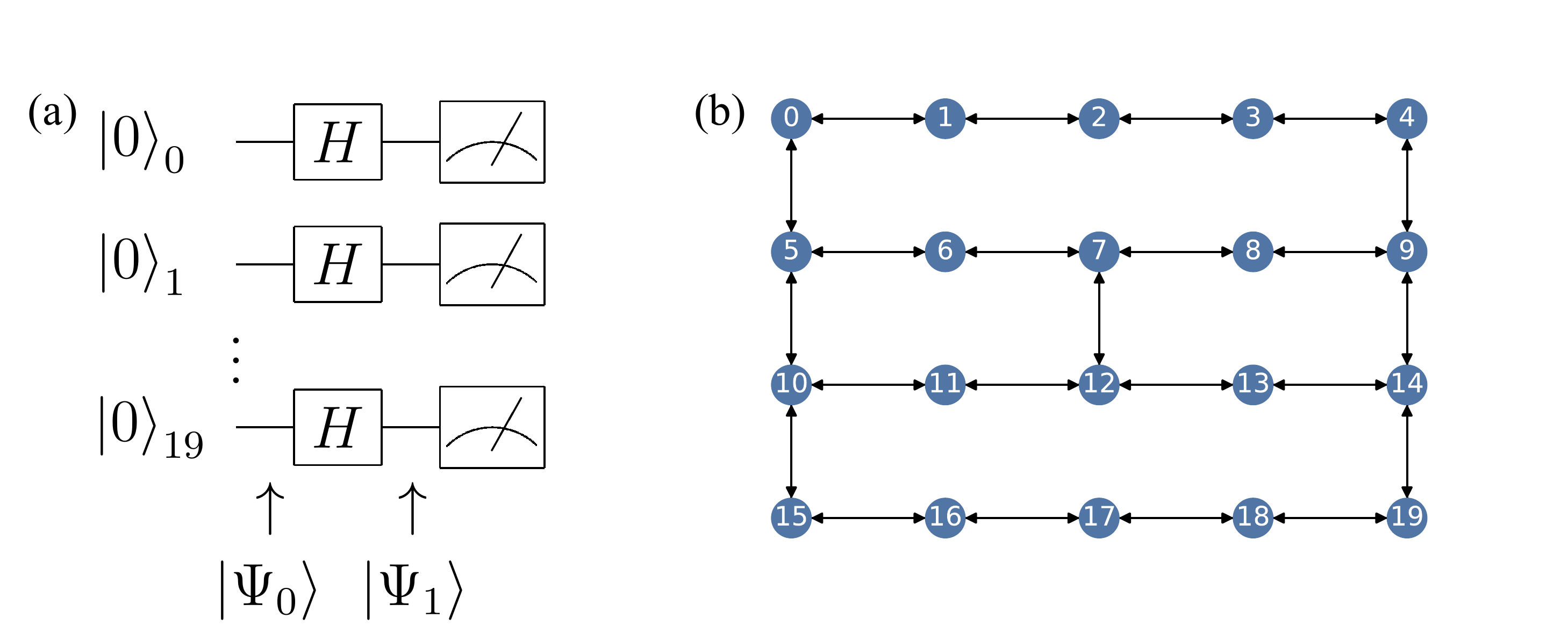}
 \caption{(a): QRNG quantum circuit using the Hadamard gate. (b): device topology of IBM 20Q Poughkeepsie provided by Qiskit.}
 \label{fig:hadamard}
 \label{fig:P_topology}
\end{figure}
In the present study, the cloud superconducting quantum computer, IBM 20Q Poughkeepsie, was used. The device was given the circuit in Fig. \ref{fig:hadamard}(a) and was repeatedly instructed to execute the circuit $8192$ times without interruption from 2019/05/09 \  11:24:27 GMT. Because the quantum computer has multiple users across the globe, interruption between jobs occur~\cite{qiskit}. 8192 is the maximum number of uninterrupted executions (shots) available. Running the circuit with 8192 shots yields a binary sequence with a length of 8192 per qubit. This process was automatically repeated across calibrations. The device goes through calibration once a day as seen in Table \ref{tab:calibration}.

As a result, $579$ samples were obtained from the IBM 20Q Poughkeepsie device. Note that each qubit produced $579$ samples, each with a length of $8192$. The samples were subjected to the eight tests from the NIST SP 800-22, which are: the frequency test, frequency within a block test, runs test, longest runs within a block test, DFT test, approximate entropy test, and the cumulative sums test (forward, backward). The p-value of each test corresponding to the respective samples was computed. For each test, the proportion of passed samples was checked. The acceptable range of the proportion of passed samples for $579$ samples under the level of significance $\alpha = 0.01$ is $> 0.977595$.
\begin{table}[ht]
\caption{The correspondence between calibration start/end time and time of job sent. All dates and times are in GMT.}
\label{tab:calibration}
\centering
\begin{tabular}{llll}\hline
\ \ \ \  \ \ \ & Start time (GMT)    & \ \  & End time (GMT) \\ \hline \hline
1 & 2019/05/08 \ 23:34:19 & & 2019/05/09 \ 05:10:24 \\
2 & 2019/05/09 \ 21:58:54 & & 2019/05/10 \ 06:23:42 \\
3 & 2019/05/10 \ 23:07:22 & & 2019/05/11 \ 02:48:12 \\
4 & 2019/05/11 \ 20:59:21 & & 2019/05/11 \ 23:33:42 \\
5 & 2019/05/12 \ 20:50:41 & & 2019/05/12 \ 23:24:58 \\ \hline
\end{tabular}
\end{table}

By constantly running the IBM 20Q Poughkeepsie device for five days, we obtained $579$ samples for each of the 20 qubits. In theory, these samples should qualify as the output of an ideal random number generator. In random number generation, the output sequences are checked for two properties: bias and patterns. When the sequences show signs of bias or patterns, the device is not in ideal condition. The same logic applies to the cloud quantum computer. We also simulated the same quantum circuit on the simulator with the obtained noise parameters such as the T1 and T2 time, the coherent error, the single-qubit error, and the readout error, all of which are updated. The simulator is referred to as the noisy simulator in the following. The noisy simulator program was also provided by IBM~\cite{qiskit}. 

In the present section, the random number output of each qubit inside the IBM 20Q Poughkeepsie device is analyzed. The qubits that are connected by arrows in Fig. \ref{fig:P_topology}(b) represent the pairs of qubits on which the controlled NOT gate can operate.  The controlled NOT gate is a two-qubit gate.

\begin{figure}[ht]
 \centering
  \includegraphics
  [height=\textwidth,angle = 90]{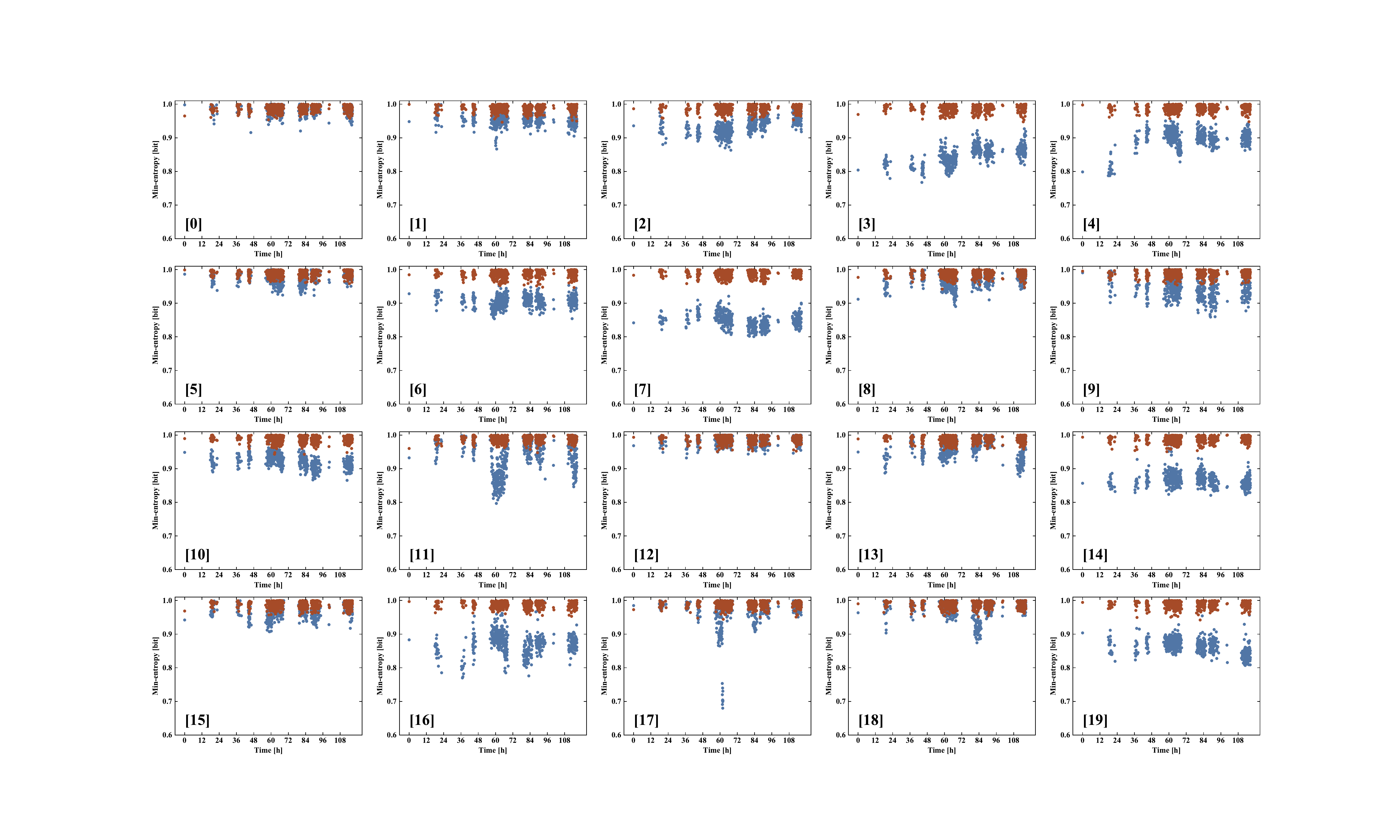}
 \caption{Min-enrtopy transition of qubit {\sf{[0]}}$\sim${\sf{[19]}}. The blue plots are the experimental results and the red plots the noisy simulation results. The figure has been rotated 90 degrees. The horizontal axis ranges from 2019/05/09 \ 11:24 GMT to 2019/05/14 \ 07:54 GMT.}
 \label{fig:min_topology}
\end{figure}
 The min-entropy, whose definition and properties are seen in the Appendix, was computed for each qubit from the 579 samples. This resulted in 579 min-entropy transition plots for 20 qubits. Figure \ref{fig:min_topology} is organized to form the topology of the IBM 20Q Poughkeepsie. 
The min-entropy takes values from 0 to 1 depending on the highest probability of the probability distribution. When the probability distribution is uniform, the min-entropy is 1. Figure \ref{fig:min_topology} shows how each qubit has a unique tendency for min-entropy. Qubit {\sf{[17]}}, for example, shows a sudden drop in min-entropy at around 60 hours. This does not occur in simulation. A sudden drop in min-entropy suggests that the measurement results can vary depending on when the cloud quantum computer executes a circuit. Overall, the noisy simulator tends to have a higher min-entropy compared to the actual device. According to Ref.~\cite{qiskit}, the readout error that IBM provides does not reflect the asymmetry between the error output 1 on the state $\ket{0}$ and the error output 0 on the state $\ket{1}$. The discrepancy between the min-entropy of the actual device and the simulator suggests that readout asymmetry exists. 
\begin{figure}[H]
 \centering
  \includegraphics[width=0.7\textwidth]{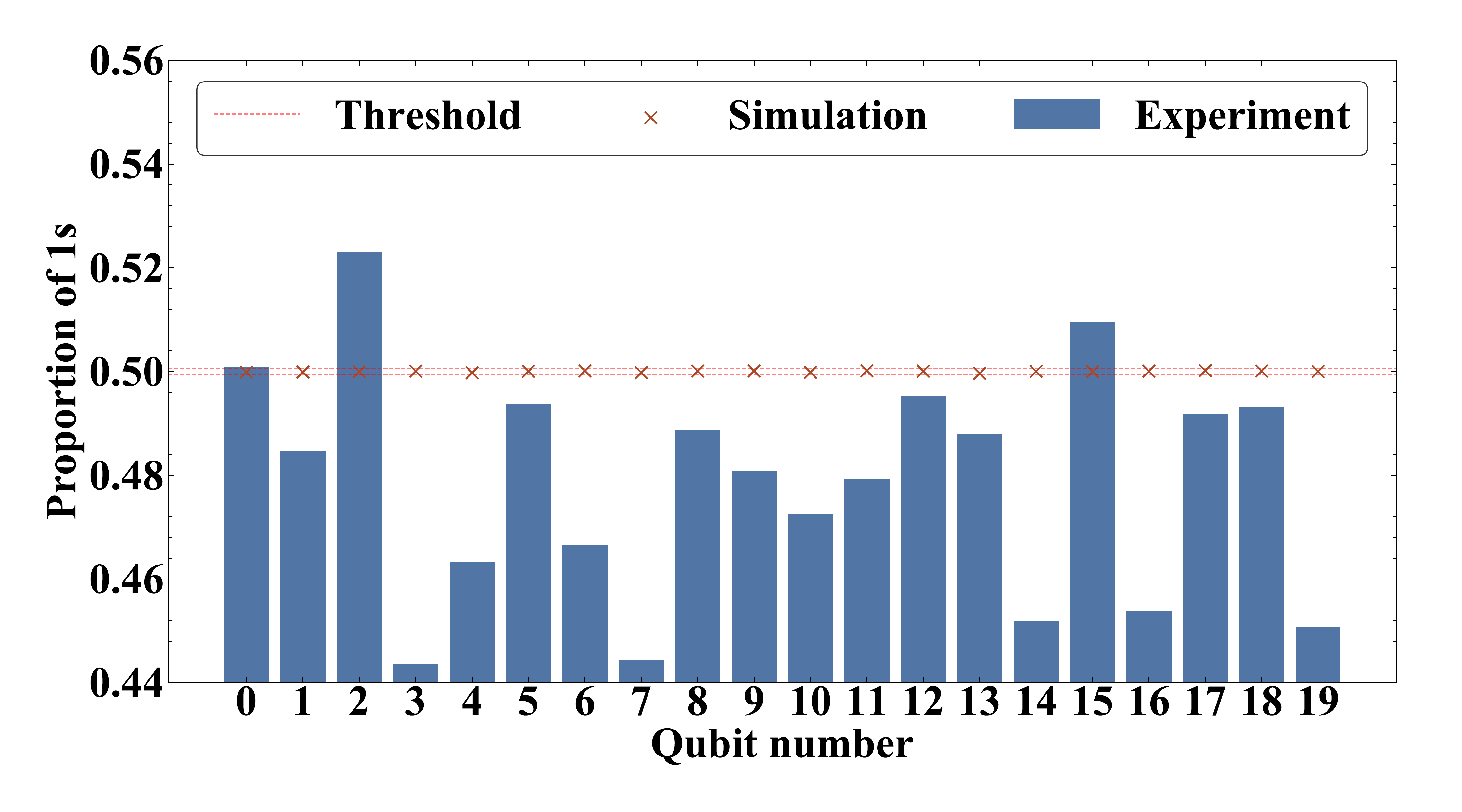}
 \caption{The proportion of 1s of qubit[0]$\sim$[19]. The acceptable range under the level of significance $\alpha = 0.01$ is between the two dotted lines. The blue bars are the experimental results and the red plots the noisy simulation results.}
 \label{fig:histogram}
\end{figure}

\begin{figure}[t]
 \centering
  \includegraphics[height=0.99 \textwidth, 
  angle = 90]{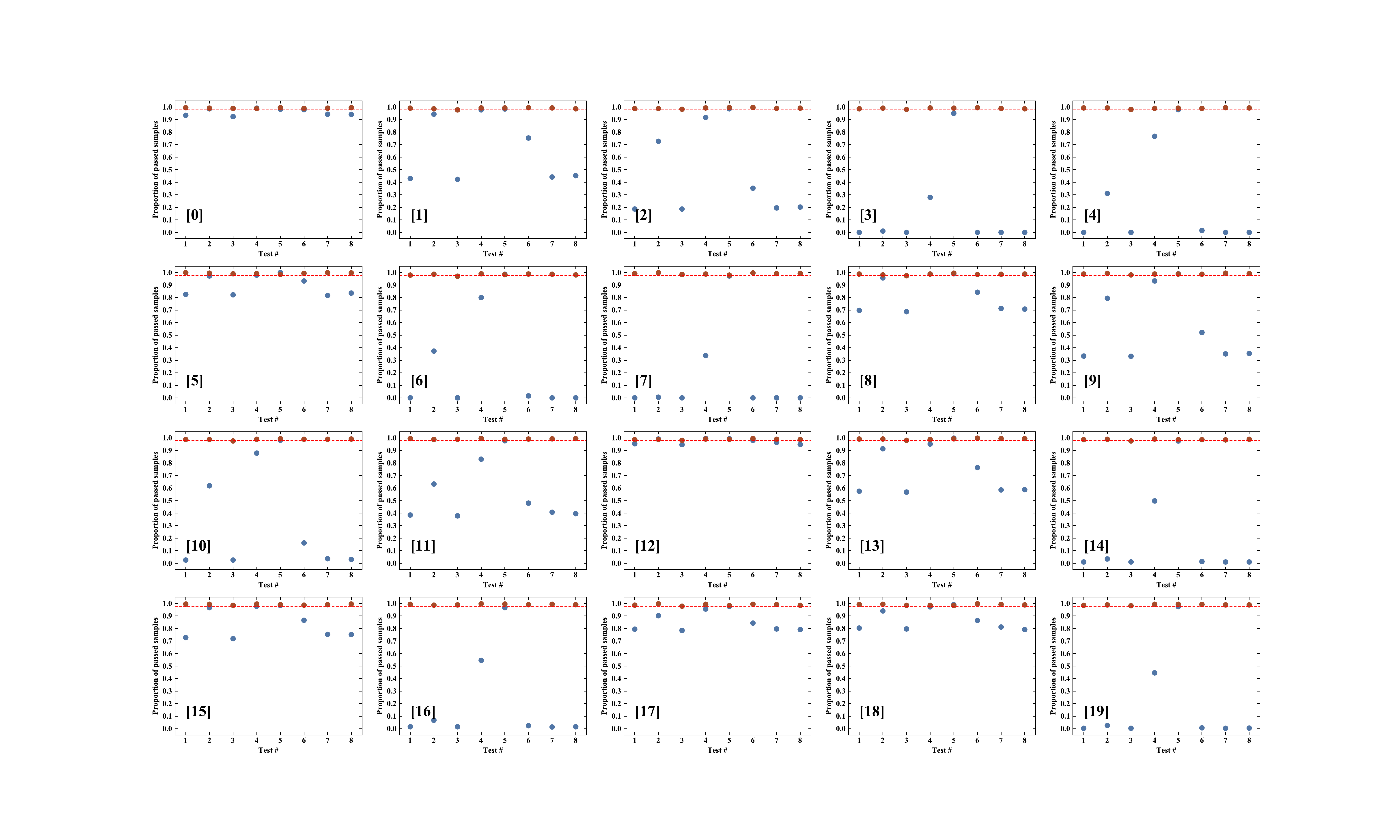}
 \caption{The proportion of passed samples for each test. The test names corresponding to the test \# can be found in Table \ref{tab:min_sample}. The acceptable range provided by the NIST is above the red line marking the proportion $0.977595$. The blue plots are the experimental results and the red plots the noisy simulation results. The figure has been rotated 90 degrees.}
 \label{fig:nist_result}
\end{figure}

Next, the samples were checked for bias. Each qubit produced 579 samples with a length of 8192, which form 4,743,168-bit sequences when chronologically connected. Figure \ref{eq:pro_of_1s} demonstrates the proportion of 1s in the entire sequence output by each qubit. Under the level of significance $\alpha = 0.01$, the proportion of 1s of a 4,743,168-bit sequence should fall between the red lines. The result is that none of the qubits produced acceptable proportions of 1s as seen in Fig. \ref{fig:histogram}. Furthermore, Fig. \ref{fig:nist_result} shows that the actual device failed to pass the eight statistical tests, which indicates that the current quantum computing device does not have the statistical properties of a uniform random number generator. 

\begin{figure}[t]
 \centering
  \includegraphics[height=0.99 \textwidth, 
  angle = 90]{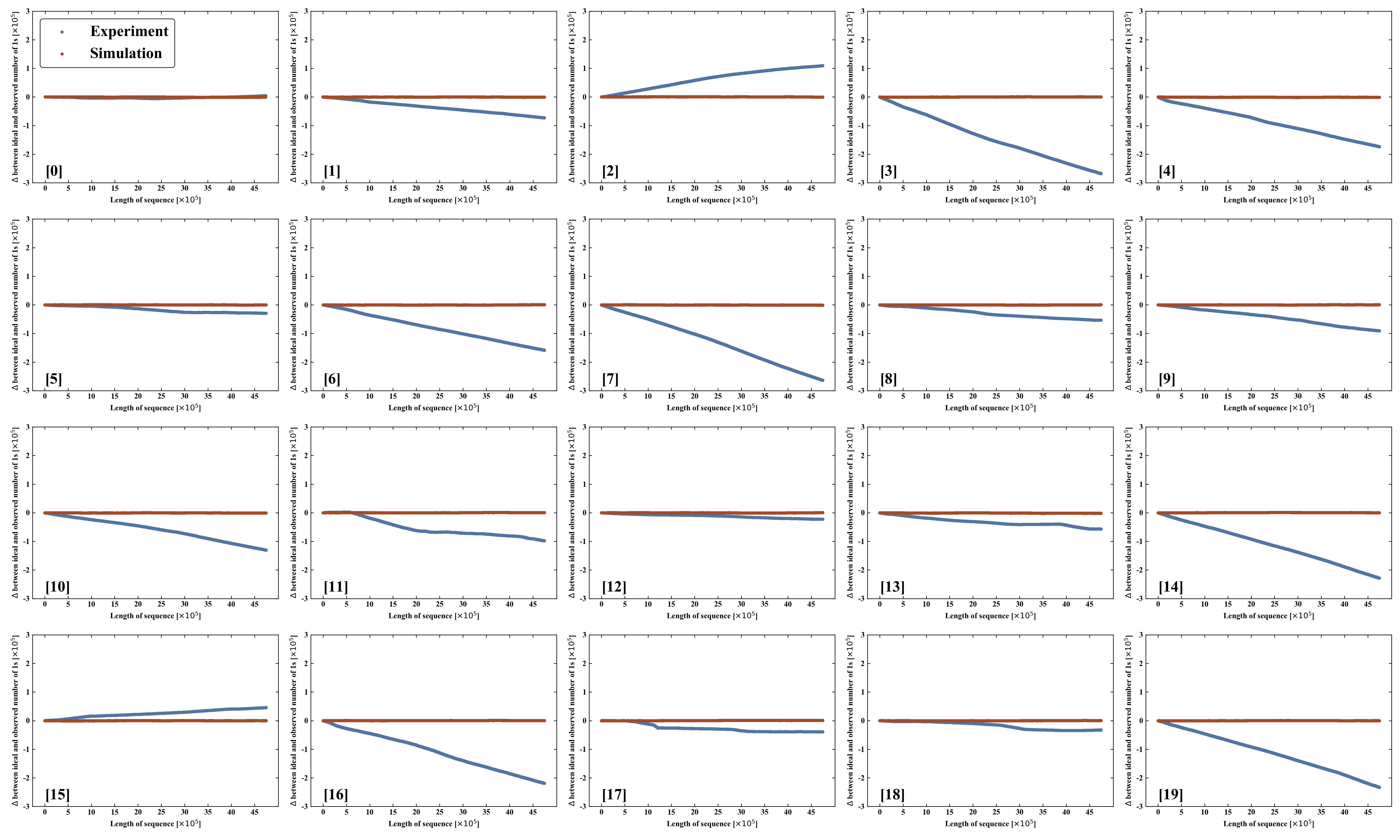}
 \caption{The difference between the ideal and observed increase in the number of 1s of qubit {\sf{[0]}}$\sim${\sf{[19]}}. The blue plots are the experimental results and the red plots the noisy simulation results. The figure has been rotated 90 degrees.}
 \label{fig:cdf}
\end{figure}

The problem with histograms as seen in Fig. \ref{fig:histogram} is that they fail to detect certain anomalies. For example, a sequence consisting of all 0s for the former half and all 1s for the latter half yields a perfect histogram. However, such a sequence is clearly not random. To compensate for this flaw, we focused on the transition of the number of 1s in the sequence. Ideally, the number of 1s in a random number sequence should always be roughly half of the sequence length. The difference between the ideal number of 1s and the observed number of 1s for the 4,743,168-bit sequence of each qubit is examined in Fig. \ref{fig:cdf}. Note that here, too, the figures are aligned topologically. Figure \ref{fig:cdf} shows the stability of each qubit in terms of the proportion of 1s in its output; a linear plot suggests that the qubit is being stably operated. While qubit[{\sf 7}] is more biased than qubit[{\sf 17}] overall, the line representing qubit[{\sf 7}] shows more stability than that of qubit[{\sf 17}]. Furthermore, the noisy simulator does not capture the trend of the qubits. Therefore, the discrepancy between the output of the actual device and the noisy simulator may not only be a result of readout asymmetry, but also time-varying parameters.

\section{Conclusion}
\label{sec:4}
We characterized the qubits in a cloud quantum computer by using statistical tests for random number generators to provide a potential indicator of the device's condition. 
The IBM 20Q Poughkeepsie device was repeatedly run for a period of five days, and 579 samples with a length of 8192 were obtained for each of the 20 qubits. For comparison, the noise parameters obtained in experiment were used to run the noisy simulator. Samples from both the actual device and the simulator were statistically analyzed for bias and patterns.
To evaluate the uniformity of each sample, the min-entropy was computed. The transition of min-entropy showed that the qubits have unique characteristics. 
We identified a sudden drop of min-entropy in qubit {\sf{[17]}}. 
The histogram of the proportion of 1s in the 4,743,168-bit sequences produced by each qubit revealed that, overall, none of the qubits produced acceptable proportions of 1s. However, we evaluated each qubit's stability from the time-series data of the proportion of 1s, and found that qubits {\sf{[0]}} and {\sf{[12]}} were relatively stable. 
Finally, eight tests from the NIST SP 800-22 were applied to the 529 samples of the 20 qubits. None of the qubits cleared the standards of the test suite. However, the test results showed that qubits {\sf{[0]}} and {\sf{[12]}} were the closest to the ideal in terms of the proportion of passed samples for each test.

As is the case with random number generators, a cloud quantum computer is a black box to its users. Therefore, users are required to decide for themselves when to use a cloud quantum computer and which qubits to choose. Statistical tests for random number generators are a potential candidate for a simple indicator of qubit condition and stability inside a cloud quantum computer.

\begin{acknowledgement}
The authors thank Hidetoshi Okutomi, Atsushi Iwasaki, Shumpei Uno and Rudy Raymond for valuable discussions. This work is partially supported by JSPS KAKENHI (Grant Nos. 17K05082 and 19H05156). The results presented in this paper were obtained in part using an IBM Q quantum computing system as part of the IBM Q Network. The views expressed are those of the authors and do not reflect the official policy or position of IBM or the IBM Q team.
\end{acknowledgement}

\section*{Appendix: Min-entropy}
\begin{figure}[H]
 \centering
  \includegraphics[width=0.70\textwidth]{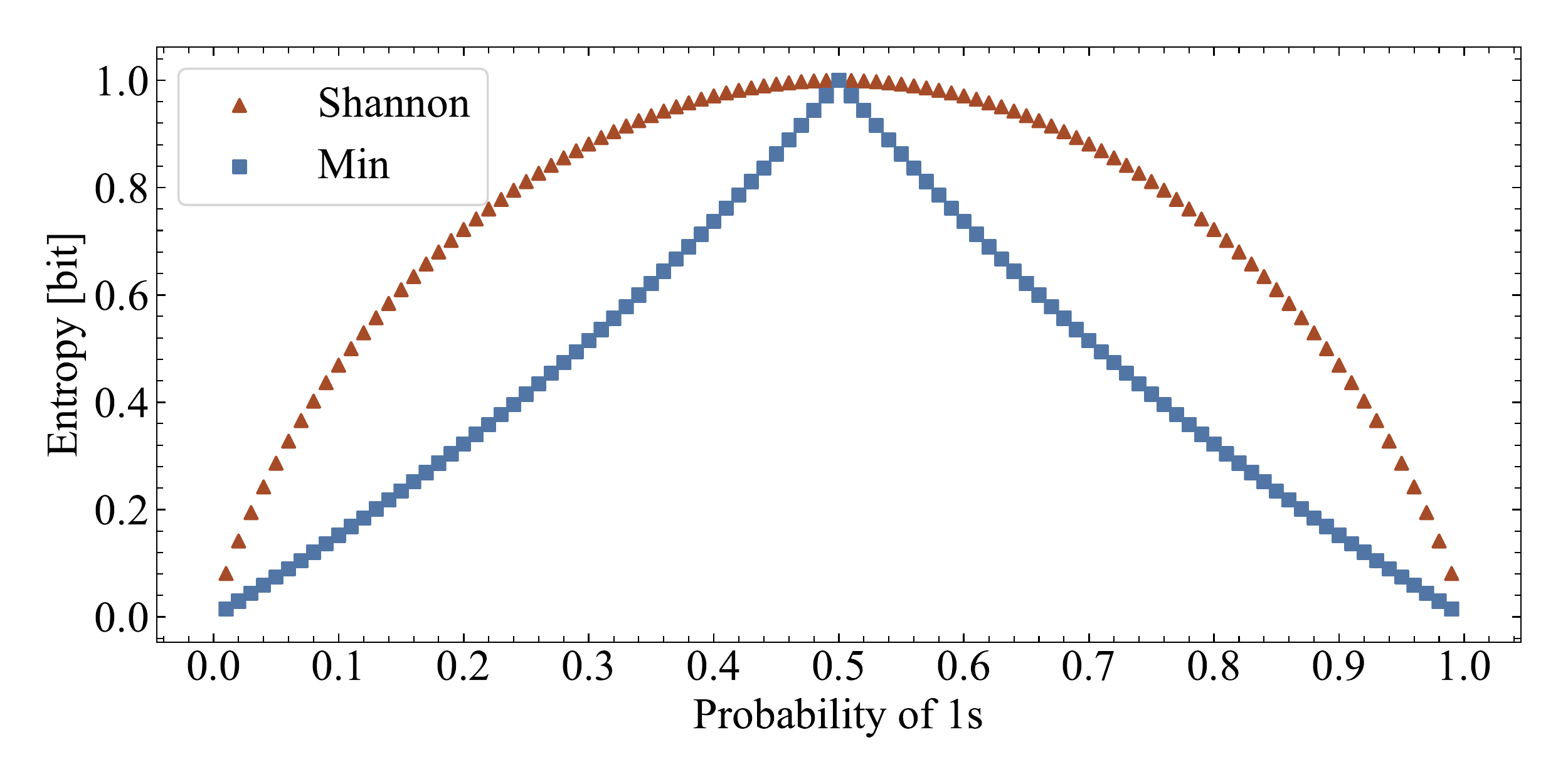}
 \caption{Relation between Shannon's entropy and min-entropy.}
 \label{fig:shannon_min}
\end{figure}

\addcontentsline{toc}{section}{Appendix}
Among various entropy measures for uniformity, the min-entropy is often used in the context of cryptography. The min-entropy for a random variable $X$ is defined as follows:
\begin{align}
    {\mathrm H_{\infty}}(X) = -\log_{2}\left( \underset{x \in \{0, 1\}}{\max}{\textrm Pr}[X = x] \right).
    \label{eq:min}
\end{align}
On the other hand, Shannon's entropy, which is also a measure for uniformity, is defined as follows:
\begin{align}
    {\textrm H_{sh}}(X) = -\underset{x \in \{0, 1\}}{\sum}
    {\textrm Pr}[X = x]\log_{2} {\textrm Pr}[X = x].
    \label{eq:shannon}
\end{align}
Both measures (\ref{eq:min}) and (\ref{eq:shannon}) take values ranging from 0 to 1 for a random variable on $\{0, 1\}$. The reason why the min-entropy is more appropriate in the context of cryptography is that it is more sensitive than Shannon's entropy. This is apparent from Fig.  \ref{fig:shannon_min}. Figure \ref{fig:shannon_min} compares the min-entropy and Shannon's entropy corresponding to the probability of $X$ yielding 1. The min-entropy provides a clearer distinction of probability distributions close to uniform than Shannon's entropy.
\noindent The min-entropy also indicates the probability that an adversary with knowledge of the probability distribution of $X$ predicts the outcome of $X$ correctly~\cite{fpga}. Here, the adversary predicts the value that appears with the highest probability. For this reason, the min-entropy considers the maximum probability of $X$.

\end{document}